\documentclass[prb,twocolumn,prb,amsmath,amsfonts]{revtex4-2}
\usepackage{graphicx}
\usepackage{epsf}
\usepackage{epstopdf}
\usepackage{graphicx}
\usepackage{tikz}
\usepackage{float}
\usepackage{amsmath,bm,upgreek}
\usepackage{nccmath}
\usepackage[mathscr]{euscript}
\usepackage{natbib}
\usepackage{hyperref}

\begin{document}
\draft
\title{Cooper pair splitter in a photonic cavity: Detection of Andreev scatterings}
\author{Bogdan R. Bu{\l}ka}
\affiliation{Institute of Molecular Physics, Polish Academy of
Sciences, ul. M. Smoluchowskiego 17, 60-179 Pozna{\'n}, Poland}

\date{Received \today \hspace{5mm} }

\begin{abstract}

We simulated the radiative response of the cavity quantum electrodynamics (QED) coupled to the double quantum dot Cooper pair splitter  and analyzed its spectral dependence to get insight into dynamics of the Cooper pair
transfers.  The model is confined to the energy subspace where two entangled electrons are transferred to two normal electrodes through the inter-dot singlet state on two proximitized quantum dots. Our research is focused on the Andreev scatterings in the subgap regime, for which the local charge susceptibility $\Pi(\omega_p)$  is derived, by means of Keldysh Green functions, in a whole bias voltage range.
In particular, in the large voltage limit the spectrum of $\Pi(\omega_p)$ is expressed by a simple analytical formula, which shows various dissipation processes related with photon-induced transitions between the Andreev bound states.

\end{abstract}

\maketitle

\section{Introduction}

In recent decades, circuit quantum electrodynamics (QED) techniques have been successfully applied to study hybrid mesoscopic systems strongly interacting with microwave photons in resonators~\cite{Cottet2017,Clerk2020,Burkard2020,Blais2021}. This approach enables us to manipulate and probe electronic degrees of freedom such as single charges and spins (in single defects and quantum dots) as well as other quantum degrees
of freedom, such as phonons (in a nanomechanical oscillator) and magnons (in a ferromagnetic spin-wave resonator). There is great interest in electronic mesoscopic circuits in cavity QED, such as  semiconductor quantum dots, nanowires
and carbon nanotubes, where the quantum coherence of single charges and spins are detectable. One can get insight into their dynamics and relaxation processes with normal metals reservoirs, ferromagnets or superconductors.
The circuit QED technique enables us to characterize exotic condensed matter states, such as the Kondo resonance or Majorana bound states (see Ref.~\cite{Burkard2020,Cottet2017} and references therein).
One can also perform coherent manipulation and single-shot
readout of the Andreev quantum dot~\cite{Janvier2015}, which is a new kind of superconducting qubit~\cite{Zazunov2003,Chtchelkatchev2003,Skoldberg2008} with the states corresponding to microscopic degrees of freedom of the
superconducting condensate.

A Cooper pair splitter (CPS), with a central superconducting electrode (as a reservoir of Cooper pairs) and two outer normal metal electrodes~\cite{Burkard2000,Lesovik2001,Recher2001,Borlin2002,Sauret2004}, was proposed the solid state setup for quantum information processing~\cite{Wendin2017}, which allows us to test Bell-inequalities by means
of current-current correlations  and to show their violation as evidence of entanglement of electrons~\cite{Chtchelkatchev2002, Samuelsson2003,Busz2017}. High efficiency of spatial entangled electrons was demonstrated
experimentally for a double-quantum-dot CPS (DQD-CPS) ~\cite{Hofstetter2009,Herrmann2010,Hofstetter2011, Schindele2012, Das2012}, also with two graphene quantum dots \cite{Tan2015} and two topologically non-trivial semiconducting nanowires \cite{Baba2018}. All of these studies
have been focused on the average currents and the zero frequency current correlations.
However, to probe dynamics of the Cooper pair splitting, one needs to use circuit QED techniques.
Such research was performed in the past decade on carbon-nanotube based DQD-CPS (CNT-DQD-CPS)~\cite{Cottet2012,CottetPRB2012,Cottet2014,Bruhat2018}, whose modeling,  besides the Cooper pair coherent splitting term, included spin-orbit interactions, inter-orbital transitions and direct inter-dot
electron hopping. It was assumed that the cavity electric field interacts with local electric dipoles  as well as induces inter-orbital transitions and
spin-flips due to spin-orbit interaction. The model describes various photon-induced excitations: singlet and triplet Cooper pairs, transitions
between them (due to spin-orbit coupling), as well as single-electron transitions. The recent experiment on CNT-DQD-CPS~\cite{Bruhat2018} used the cavity QED as a spectroscopic probe and demonstrated Cooper-pair-assisted cotunneling between the quantum dots, in equilibrium conditions.

We want to revisit the DQD-CPS model in its simplified form,
 where two entangled electrons are transferred through the inter-dot singlet state on two proximized QDs into two normal electrodes.
 Using the Keldysh Green function technique we can get insight into quantum coherence processes in electronic transport and dynamics of the Cooper pair
transfers through various Andreev bound states (ABS) in non-equilibrium conditions, for a whole bias voltage range~\cite{Chevallier2011,Bulka2021}.
These features will be analyzed quantitatively, studying the radiative response of a microwave cavity, in terms of the local charge susceptibility of the DQD-CPS for realistic model parameters (close to recent experiments).
Different dissipation processes of split Cooper pairs will be extracted by means of spectral decomposition of  the charge susceptibility.

\section{Model description and derivation of cavity response}
\label{model}

\begin{figure}
\centering
\includegraphics[width=.92\linewidth,clip]{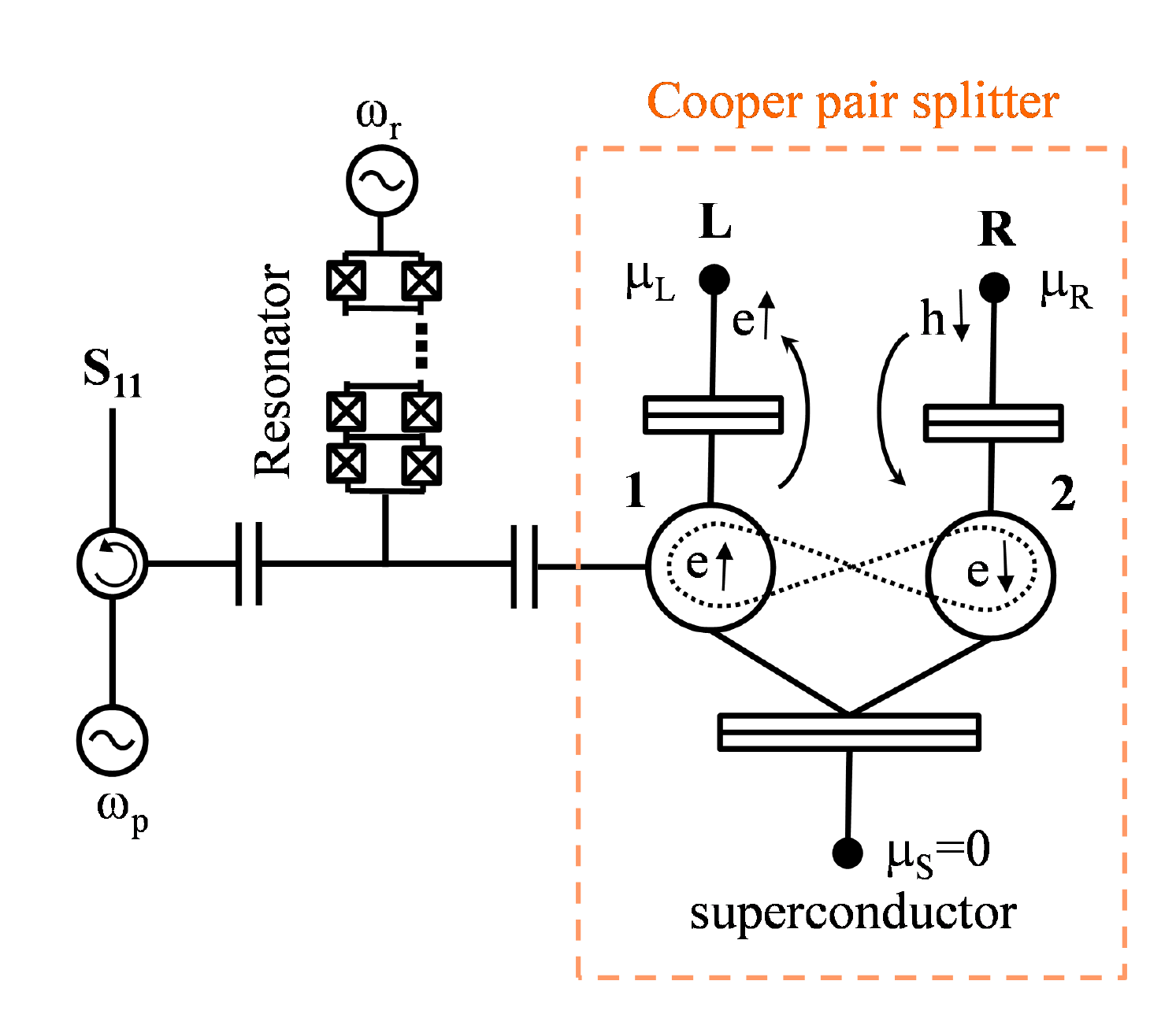}
\caption{Schematic presentation of the Cooper pair splitter (CPS), with two quantum dots ($1,2$) coupled to the normal ($L, R$) electrodes and strongly coupled to the superconductor ($S$) as a reservoir of Cooper pairs.
Charge transport in CPS  is due to perfect crossed Andreev reflections (CAR) when an electron (e) is injected to the normal electrode and a hole (h) with an opposite spin is simultaneously ejected from the second metallic
electrode. We assume that CPS is coupled to a SQUID array resonator and a microwave detection system for reflectance measurements (adapted from Refs.~[\onlinecite{Stockklauser2015,Stockklauser2017,Kratochwil2021,Scarlino2019}]). }
\label{fig1}\end{figure}

We assume that our mesoscopic system is embedded in a microcavity and their interaction is described in the framework a semiclassical linear response
approach, the input-output theory~\cite{Walls2008,Cottet2017,Burkard2020,Bruhat2016}. For the single-sided resonator  the reflection coefficient can be derived as~\cite{Kratochwil2021}
\begin{align}
S_{11} \equiv\frac{a_{out}}{a_{in}}
=-\frac{\omega_p - \omega_r + \imath (\kappa_{int} - \kappa_{ext})/2- \Pi(\omega_{p})}{\omega_p - \omega_r  + \imath (\kappa_{int} +\kappa_{ext})/2-\Pi(\omega_{p})},
\end{align}
where  $\omega_r-\omega_p$ is the detuning of the resonator frequency from the probe frequency $\omega_p$ (the cavity drive frequency), $\kappa_{int}$ and $\kappa_{ext}$ denote internal and external resonator dissipation rates.
Here, a key quantity of interest is $\Pi(\omega_{p})=\sum_{i,j} g_i g_j \chi_{i,j}(\omega_{p})$ -- the Fourier transform of the charge susceptibility: $\Pi(t-t')=-\imath\theta(t-t')\sum_{i,j} g_i g_j\langle[n_i(t),n_j(t')]\rangle_{g=0}$, with the average performed over the electronic system decoupled from the cavity.
We have assumed that the light–-matter interaction is
well approximated by dipolar coupling with a local charge, described by  $H_{cav-dip}=\sum_i g_i n_i(a^\dag+a)$, where $a^\dag$ denotes the cavity photon creation operator, $n_i$ is the local charge operator and $g_i$ is the local coupling strength \cite{Cottet2017,Cottet2015}. The mutual capacitive coupling between the two dots
is disregarded.
 The two-particle averages are decoupled by means of Wick’s theorem to products of single-particle averages, which are then expressed by the Keldysh Green
functions. The result is~\cite{Bruhat2016,Cottet2017,Cottet2020}
\begin{align}\label{chi}
\chi^*_{i,j}(\omega_{p})=-\imath\int \frac{dE}{2\pi} \text{Tr}&\{[\tau_i G^r(E+\hbar\omega_p)\tau_j+\nonumber\\
&\tau_j G^a(E-\hbar\omega_p)\tau_i]G^<(E)\},
\end{align}
where $G^{r,a,<}$ denote the retarded, the advanced and the lesser Green functions, $\tau_i=\text{diag}(1,-1)$ is the matrix describing the structure of the photon-particle coupling in the Nambu (electron-hole) space.
This approach takes into account coherent processes inside the nanosystem as well as coherent coupling with electrodes. It works very well  for a single
quantum dot system~\cite{Bruhat2016}, recovers the susceptibility in double dots derived within the master equation approach when dissipation is due sequential tunneling~\cite{Cottet2017}.

Let us specify our Cooper pair splitter; it consists of two quantum dots (DQD), where each QD is coupled to the normal $L$ or $R$ electrode and both are coupled the superconductor $S$ - see Fig.1. The corresponding Hamiltonian is
\begin{align}
\label{eq:ham}
H_{CPS} = & \sum_{\alpha,k}\Psi^{\dag}_{\alpha k}(\epsilon_{\alpha k}\sigma_z+\Delta_{\alpha}\sigma_x)\Psi_{\alpha k}+ \sum_{i} d^{\dag}_{i}\varepsilon_{i}\sigma_z d_{i}\nonumber\\
&+ \sum_{\alpha,k,i}\big(\Psi^{\dag}_{\alpha k}t_{\alpha i}\sigma_z d_{i}+H.c.\big) \; ,
\end{align}
where the first term describes the electrodes $\{\alpha=L,R,S\}$ in  Nambu notation $\Psi^{\dag}_{\alpha k}=(c^{\dag}_{\alpha k\uparrow},c_{\alpha \bar{k}\downarrow})$, $\bar{k}=-k$,
$\sigma_z$, $\sigma_x$ are the Pauli matrices, $\epsilon_{\alpha k}$ and $\Delta_{\alpha}$ denotes the electron energy and the superconducting gap, with $\Delta_{L,R}=0$ for the normal electrodes.  The second term corresponds to the QDs, $\{i=1,2\}$, with a single level $\varepsilon_{i}$, where $d^{\dag}_{i}=(c^{\dag}_{i\uparrow},c_{i\downarrow})$ is a spinor in Nambu notation for the local QD operator. The last term describes
coupling of DQD with the electrodes, where $t_{\alpha i}$ denotes  the electron hopping between the $\alpha$ electrode and the $i$-th QD (as shown in Fig.1).

In the DQD system several many-electron states with different charge and spin configurations can occur. For the proximized system with two electrons,  the lowest state is the interdot singlet pairing, whereas the intradot pairing is much higher in an energy scale due to a large intradot Coulomb repulsion. For sufficiently low probe signals one can confine considerations to the lowest subspace with the interdot singlet, $\langle c^{\dag}_{1\uparrow}c^{\dag}_{2\downarrow}-c^{\dag}_{1\downarrow}c^{\dag}_{2\uparrow}\rangle \neq 0$.
To calculate the charge density response, $\chi^*_{i,j}(\omega_{p})$, we use the Keldysh Green function method, following Ref.[\onlinecite{Chevallier2011}] and [\onlinecite{Bulka2021}].
Since our interest is in the Andreev scatterings, therefore the calculations are performed in the subgap regime $|E|<\Delta_{S}$  and the limit $\Delta_{S} \rightarrow \infty$, in which the Green function has the self-energy \cite{Chevallier2011,Bulka2021}:
\begin{eqnarray}\label{eq:S-S}
\hat{\Sigma}_{S}=\frac{\gamma_{S}}{2}\left[\begin{array}{cccc}
0&0&1&0\\
0&0&0&-1\\
1&
 0&0&0\\
 0&-1&0&0
\end{array}\right],
\end{eqnarray}
where $\gamma_{S}=\pi \rho_S t_{S1}t_{S2}$ is the interdot exchange electron-hole coupling,  which describes the Cooper pair coherent splitting, and $\rho_S$ denotes the density of states in the S-electrode in
the normal state. In this way the S-electrode is integrated out and the
system consists of two proximized QDs (with the inter-dot singlet) coupled to two normal electrodes.  The Keldysh Green function matrix is expressed as a product of two components
\begin{eqnarray}\label{eq:greentot}
\hat{G}_{L2QDR}=\hat{G}_{\text{e}\uparrow,\text{h}\downarrow}\otimes \hat{G}_{\text{h}\downarrow,\text{e}\uparrow},
\end{eqnarray}
where
\begin{widetext}
\begin{gather}
\hat{G}_{\text{e}\uparrow,\text{h}\downarrow}\equiv  \left[\begin{array}{cccc}
  \hat{G}_{L\text{e}\uparrow,L\text{e}\uparrow} & \hat{G}_{L\text{e}\uparrow,1\text{e}\uparrow} & \hat{G}_{L\text{e}\uparrow,2\text{h}\downarrow}&\hat{G}_{L\text{e}\uparrow,R\text{h}\downarrow} \\
  \hat{G}_{1\text{e}\uparrow,L\text{e}\uparrow} & \hat{G}_{1\text{e}\uparrow,1\text{e}\uparrow} & \hat{G}_{1\text{e}\uparrow,2\text{h}\downarrow}&\hat{G}_{1\text{e}\uparrow,R\text{h}\downarrow} \\
 \hat{G}_{2\text{h}\downarrow,L\text{e}\uparrow} & \hat{G}_{2\text{h}\downarrow,1\text{e}\uparrow} & \hat{G}_{2\text{h}\downarrow,2\text{h}\downarrow}&\hat{G}_{2\text{h}\downarrow,R\text{h}\downarrow} \\
  \hat{G}_{R\text{h}\downarrow,L\text{e}\uparrow} & \hat{G}_{R\text{h}\downarrow,1\text{e}\uparrow} & \hat{G}_{R\text{h}\downarrow,2\text{h}\downarrow}&\hat{G}_{R\text{h}\downarrow,R\text{h}\downarrow}
\end{array}\right]
=\left[\begin{array}{cccccccc}
  w_{L,11}^{--} & w_{L,11}^{-+} & t_{L1}&0 &0 &0 &0&0 \\
  w_{L,11}^{+-} &  w_{L,11}^{++} &  0&-t_{L1} &0 &0 &0&0 \\
t_{L1}& 0 & z_{1\text{e}}& 0&\gamma_{S}/2 &0 &0 &0 \\
   0 &  -t_{L1}& 0& -z_{1\text{e}}&0 &-\gamma_{S}/2 &0 &0 \\
0 &0 &\gamma_{S}/2 &0 &  z_{2\text{h}} & 0 & -t_{R2}&0   \\
0 &0 &0 &-\gamma_{S}/2 & 0 &   -z_{2\text{h}} & 0 & t_{R1}  \\
0 &0 &0 &0 &  - t_{R2}& 0 & w_{R,22}^{--}& w_{R,22}^{-+}  \\
0 &0 &0 &0 &    0 & t_{R2}&   w_{R,22}^{+-} & w_{R,22}^{++}
\end{array}\right]^{-1}.\label{eq:NEGF}
\end{gather}
Here, the Keldysh notation is used for the Green functions. The inverse elements of the Green function $\hat{G}_{L\text{e}\uparrow,L\text{e}\uparrow}$ (for electrons (e) in the L electrode) and $\hat{G}_{R\text{h}\downarrow,R\text{h}\downarrow}$ (for holes (h) in the R electrode) are: $w_{L,11}^{--}= w_{L,11}^{++}=-2 \imath \rho_L (f_{L\text{e}}-1/2)$, $w_{L,11}^{-+}= 2 \imath \rho_L f_{L\text{e}}$, $w_{L,11}^{+-}= -2 \imath \rho_L (1-f_{L\text{e}})$  and $w_{R,22}^{--}= w_{R,22}^{++}=-2 \imath \rho_R (f_{R\text{h}}-1/2)$, $w_{R,22}^{-+}= 2 \imath \rho_R f_{R\text{h}}$, $w_{R,22}^{+-}= -2 \imath \rho_R (1-f_{R\text{h}})$.
$f_{\alpha \text{e}}=\{\exp[(E-\mu_{\alpha})/k_BT]+1\}^{-1}$ and  $f_{\alpha \text{h}}=\{\exp[(E+\mu_{\alpha})/k_BT]+1\}^{-1}$ are the Fermi distribution functions for electrons  and holes  in the $\alpha$ electrode with the chemical potential $\mu_{\alpha}$, at the temperature $T$, with $k_B$ as the Boltzmann constant. The chemical potential in the superconductor is taken to be $\mu_S=0$. We also denoted $z_{1\text{e}}=E-\varepsilon_1$ and $z_{2\text{h}}=E+\varepsilon_2$.

The retarded, the advanced and the lesser Green functions are derived using the relations: $G^r=G^{--}-G^{-+}$, $G^a=G^{--}-G^{+-}$ and $G^<=G^{-+}$. In the next step, these functions are inserted into Eq.(\ref{chi}) and the local charge susceptibility is expressed as:
\begin{align}\label{chi1e1e}
\chi^*_{1\text{e},1\text{e}}(\omega_p)=\int \frac{dE}{2\pi} \frac{8 [f_{L\text{e}} \gamma_L (4 z_{2\text{h}}^2+\gamma_R^2)+f_{R\text{h}} \gamma_R \gamma_S^2] }{[(2 z_{1\text{e}}+\imath \gamma_L) (2 z_{2\text{h}}+\imath \gamma_R)-\gamma_S^2][(2 z_{1\text{e}}-\imath \gamma_L)(2 z_{2\text{h}}-\imath \gamma_R)-\gamma_S^2]}
\nonumber\\
\times\Bigg[\frac{2 z_{2\text{h}}^+ + \imath \gamma_R}{(2 z_{1\text{e}}^+ +\imath \gamma_L) (2 z_{2\text{h}}^+ +\imath \gamma_R)-\gamma_S^2}+\frac{2 z_{2\text{h}}^- - \imath \gamma_R}{(2 z_{1\text{e}}^- -\imath \gamma_L) (2 z_{2\text{h}}^- -\imath \gamma_R)-\gamma_S^2 }\Bigg],
\end{align}
\end{widetext}
where $z_{1\text{e}}^{\pm}=E\pm\hbar\omega_p-\varepsilon_1$, $z_{2\text{h}}^{\pm}=E\pm\hbar\omega_p+\varepsilon_2$, $\hbar\omega_p$ is the energy of a photon (with $\hbar=h/2\pi$, $h$ being the Planck constant), $\gamma_L=\pi \rho_L t_{L1}^2$ and  $\gamma_R=\pi \rho_R t_{R2}^2$.
The poles of the integrand show positions of the pair of the ABS:   $E_{\pm}^{\text{eh}}=(\delta\pm\Omega)/2$, where $\Omega=\sqrt{\varepsilon^2+\gamma_S^2}$ is the separation between the ABS, $\varepsilon=(\varepsilon_1+\varepsilon_2)/2$ and $\delta=(\varepsilon_1-\varepsilon_2)/2$ denotes the level detuning.
Similarly, one gets the charge susceptibility for holes at the 1-st QD, $\chi^*_{1\text{h},1\text{h}}(\omega_p)$, exchanging the electron and hole channels
$\{\text{e}\leftrightarrow \text{h}\}$ in Eq.(\ref{chi1e1e}). In this case the ABS are at $E_{\pm}^{\text{he}}=(-\delta\pm\Omega)/2$.

The integral in Eq.(\ref{chi1e1e}) can be calculated
numerically or analytically (using partial fraction decomposition of the spectral functions). In general, the analytical results are rather lengthy for
presentation, and therefore, we will  present and discuss plots instead.

\section{The results}

\begin{figure}
\centering
\includegraphics[width=.94\linewidth,clip]{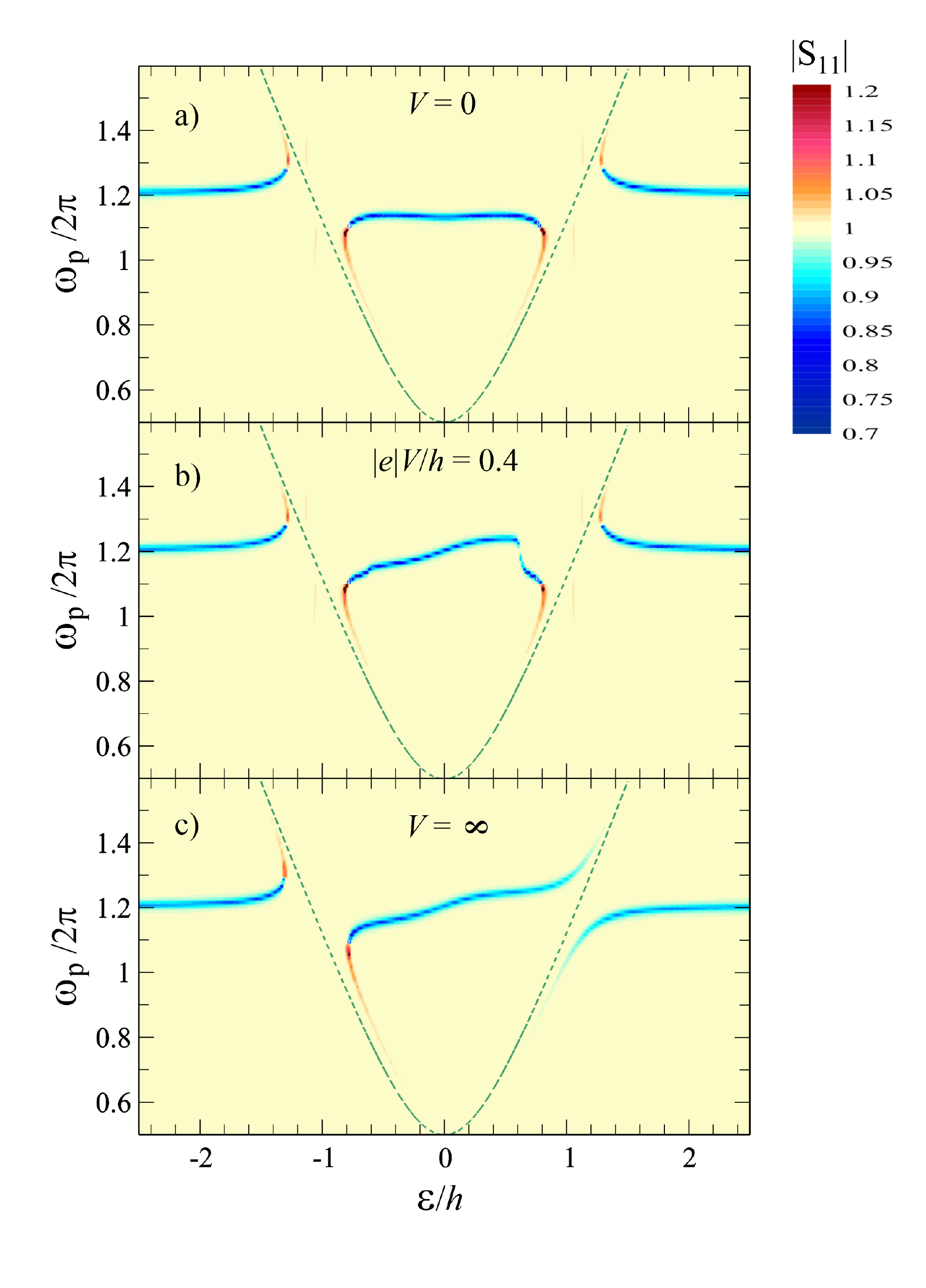}
\caption{Resonator reflectance $|S_{11}|$ plotted as a function of ($\varepsilon,\omega_p)$ for a bias voltage $|e|V/h=0$, 0.4, $\infty$,  calculated at temperature $T=0$  and the level detuning
$\delta=0$. All quantities and the parameters are in units of GHz:  $\omega_r/2\pi=1.2$, $\kappa_{int}/2\pi$ = 0.014, $\kappa_{ext}/2\pi$ = 0.001, $g_1/h$ = 0.4, $g_2=0$, $\gamma_S/h$ = 0.5
 and a symmetric QD coupling with the electrodes $\gamma_L/2\pi=\gamma_R/2\pi=\gamma/2\pi$ = 0.01.  The green dashed curve presents the dispersion relation of the DQD-CPS qubit: $\hbar\omega_p=\Omega\equiv\sqrt{\varepsilon^2+\gamma_S^2}$.
 }
\label{mapa}
\end{figure}

Let us now analyze the charge susceptibility and how its features can be seen in a reflectance measurement. Fig.\ref{mapa} shows simulation of the resonator reflectance spectrum $|S_{11}|$ as
a function of the probe frequency $\omega_p$ and the position $\varepsilon$, with respect to $\mu_S=0$.
The calculations have been performed at temperature $T=0$, for a bias voltage applied in the splitter configuration, $\mu_L=\mu_R=-|e|V \leq 0$, and a strong and asymmetric charge-photon coupling,  $g_1/h=0.4$, $g_2=0$.
 In our analysis, we follow the experimental papers~[\onlinecite{Stockklauser2015,Stockklauser2017,Kratochwil2021,Scarlino2019}], and express  all of the system parameters in units of GHz. The resonator parameters are taken as: $\omega_r/2\pi=1.2$, $\kappa_{int}/2\pi$ = 0.014, $\kappa_{ext}/2\pi$ = 0.001 (close to the recent experiment~\cite{Kratochwil2021}) and the Cooper pair coherent
 splitting parameter as $\gamma_S/h$ = 0.5 (close to $\gamma_S^{\text{exp}}/h$ = 0.4 determined on CNT-DQD-CPS~\cite{Bruhat2018}).
 As can be seen, the cavity photons and DQD-CPS qubit are at resonance when $\hbar\omega_r=\sqrt{\varepsilon^2+\gamma_S^2}$. For the chosen parameters such resonant value of $\varepsilon$ happens for $\varepsilon_r/h = \pm 1.09087$.
 Notice that $|S_{11}|>1$ in
 some regions (in red) corresponding to photon gain (a similar feature was observed in CNT-DQD-CPS~\cite{Bruhat2018}). The shape of the resonance
 depends on the voltage applied to the normal electrodes. At $V=0$ the
 reflectance $|S_{11}|$ is symmetric, but with an increase of the voltage it becomes asymmetric, because electron transport changes dissipation in the system.
Our main purpose is to show that the cavity spectroscopy can be used to studies charge dynamics and related dissipation processes in the mesoscopic system.  Let us consider this issue in greater detail.

In the large voltage limit, $V\rightarrow \infty$, one can get simple analytical results,  integrating (\ref{chi1e1e}) by means of the residue theorem. For  the symmetric coupling to the electrodes, $\gamma_L=\gamma_R=\gamma$, the charge susceptibility can be expressed as
\begin{align}\label{chi1e1einf}
\chi_{1\text{e}1\text{e}}^*(\omega_p)=& \frac{\varepsilon (\hbar\omega_p + 2 \imath \gamma)\gamma_S^2}{ 2 (\gamma^2 + \Omega^2)(\hbar\omega_p + \imath \gamma) [(\hbar\omega_p +
      \imath \gamma)^2 - \Omega^2] }\nonumber\\
      =& - \frac{\imath  \varepsilon \gamma_S^2}{2 \Omega^2( \gamma^2 + \Omega^2)(\hbar\omega_p +
      \imath  \gamma)   }\nonumber\\ & + \frac{ \varepsilon \gamma_S^2(\Omega+\imath\gamma)}{4 \Omega^2 ( \gamma^2 + \Omega^2) (\hbar\omega_p - \Omega+ \imath  \gamma ) }\nonumber\\ &+ \frac{ \varepsilon \gamma_S^2(-\Omega+\imath\gamma)}{4 \Omega^2 ( \gamma^2+ \Omega^2)  (\hbar\omega_p + \Omega+ \imath  \gamma )}.
      \end{align}
 This quantity describes local electron fluctuations at the first QD caused only by hole transfers through the R-electrode (the electron transfers from the L-electrode are prohibited).
Similarly, one can derive the susceptibility for holes, which is $\chi^*_{1\text{h},1\text{h}}(\omega_p)=\chi^*_{1\text{e},1\text{e}}(\omega_p)$. Notice that in this limit $\chi_{1\text{e},1\text{e}}$ is independent of the level detuning $\delta$, because both of the ABS equally
participate in transport and the current $I^{\infty}_L=(e/2\hbar)\gamma\gamma_S^2/[(\gamma^2+\Omega^2)]$ is independent of $\delta$ as well.
Above we have  performed also  spectral decomposition of $\chi^*_{1\text{e},1\text{e}}(\omega_p)$ to find relaxators which describe dissipation processes in the CPS system. Its first term (the second row in Eq.(\ref{chi1e1einf})) corresponds to intra-level charge
fluctuations, whereas the second and third term correspond inter-level fluctuations with absorption and emission of photons, respectively. The
relaxation rate is $1/\tau_{relax}=\gamma$, the same for all dissipation processes. These features are seen in  Fig.\ref{mapa}c, quite pronounced at the resonances,  $\varepsilon_r/h=\pm 1.09087$, and a small fold at the center, $\varepsilon=0$.
Similar charge dynamics was seen in  the cross and the auto-current correlations, with two resonant side dips related with absorption and
emission of photons (see Eqs.(32)-(33) in Ref.[\onlinecite{Bulka2021}]).

\begin{figure}
\centering
\includegraphics[width=1 \linewidth,clip]{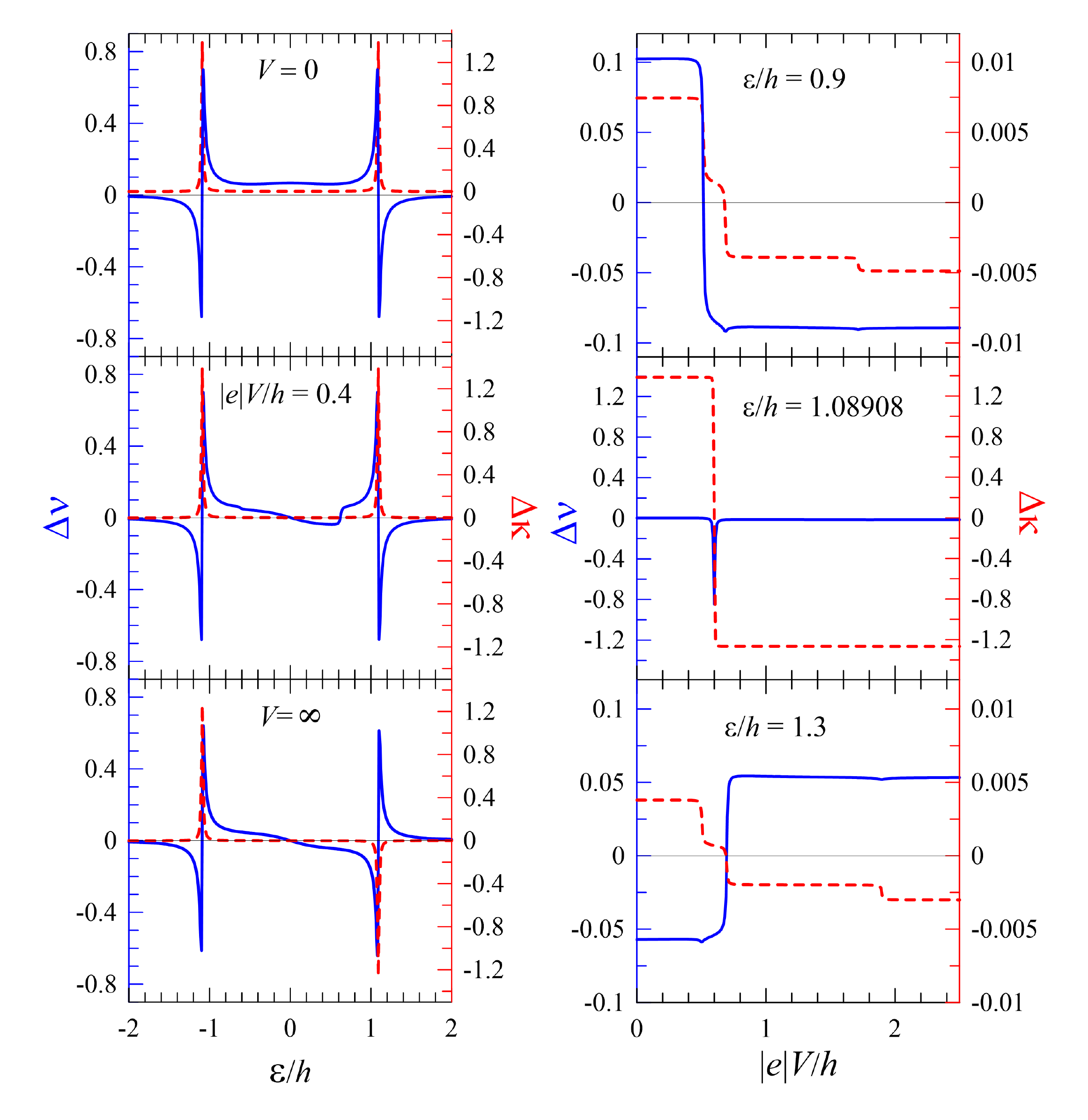}
\caption{Left column: Plots for the frequency shift $\Delta\nu\equiv \text{Re}[\Pi(\omega_r)]/h$ (the blue curve and the left vertical axis) and the  broadening $\Delta\kappa\equiv \text{Im}[\Pi(\omega_r)]/h$  (the red dashed curve and the right vertical axis)
  as a function of $\varepsilon$ for the resonant frequency $\omega_r/2\pi=\omega_p/2\pi=$ 1.2 and various $|e|V/h=$ 0, 0.4, $\infty$.
 Right column: Plots of $\Delta\nu$ and  $\Delta\kappa$
  as a function of $|e|V/h$ for  various $\varepsilon/h=$ 0.9 , 1.0908, 1.3.  The other parameters are the same as in Fig.2.
 }
\label{fig3ab}
\end{figure}

Let us study charge dynamics for a finite bias voltage. The left column in Fig.~\ref{fig3ab} presents a cavity frequency shift, $\Delta\nu= \text{Re}[\Pi(\omega_r)]/h$,  and the  linewidth broadening, $\Delta\kappa\equiv \text{Im} [\Pi(\omega_r)]/h$ as a function of $\varepsilon/h$.
In the top panel the cavity response for the nanocircuit at equilibrium is presented, and its modifications caused by electron transport are shown in the two subsequent panels. At a low bias voltage the current is
small, because the ABS lie outside the transport window. For a larger voltage, $|eV|>\Omega/2=\sqrt{\varepsilon^2+\gamma_S^2}/2$,  the central ABS states become participate in transport. For the considered case (with $|e|V/h=0.4$)  the active
transport window is for $|\varepsilon|<0.6245$.
Large charge fluctuations are seen close to the resonance points  $\varepsilon_r/h=\pm 1.09087$.

The right column in Fig.~\ref{fig3ab} presents the voltage dependence of $\Delta \nu$ and $\Delta\kappa$, for $\varepsilon$ close to the resonance value $\varepsilon_r$. Notice the different scales of the axes in the middle panel,
when large charge fluctuations are present.  The curves for $\Delta\kappa$ show steps at $|e|V_1=\Omega/2$, $|e|V_2=\hbar\omega_r-\Omega/2$ and
$|e|V_3=\hbar\omega_r+\Omega/2$, which are related with activation of dissipation processes (without and with photons through the ABS). One can see also small kinks in $\Delta \nu$ at these voltages.

\begin{figure}
\centering
\includegraphics[width=.94\linewidth,clip]{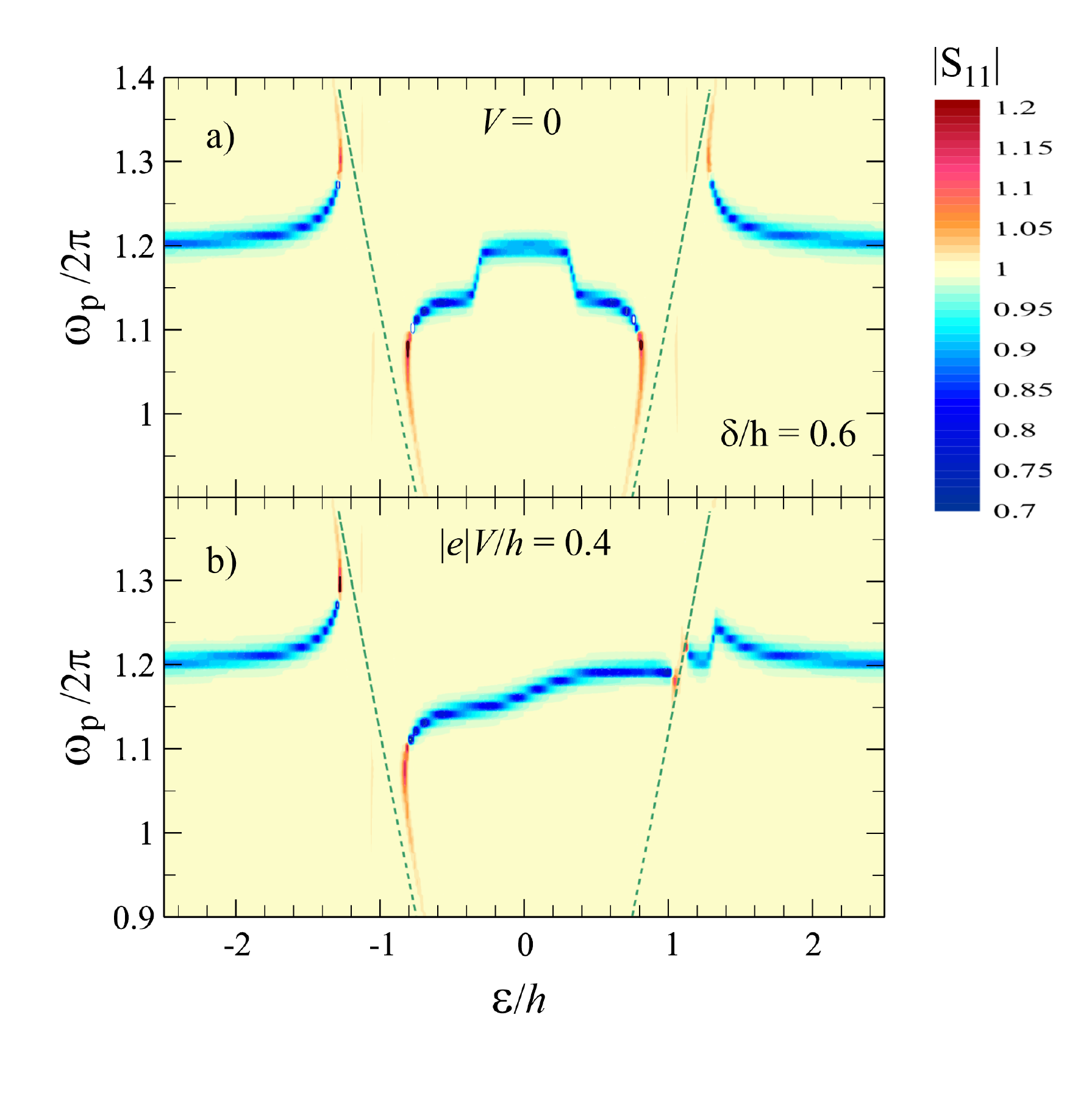}
\caption{Resonator reflectance $|S_{11}|$ plotted as a function of ($\varepsilon,\omega_p)$ for $\delta/h=0.6$ and for bias voltages $|e|V/h=0$ (top) and 0.4 (bottom).  The other parameters are the same as in Fig.2.
 }
\label{mapad6}
\end{figure}

Our consideration concerned so far the case $\delta=0$, i.e. when the dot levels $\varepsilon_1=\varepsilon_2=\varepsilon$. It is known that the level detuning $\delta$ destroys entanglement of split Cooper pairs and lowers
correlations between the split electron currents~\cite{Chevallier2011,Michalek2021}.
We have performed calculations to see a role of the detuning on the cavity response. The resonator reflectance spectrum $|S_{11}|$ is presented in Fig.\ref{mapad6} for $|e|V/h=$ 0 and 0.4, for the voltage range where one can
 expect a pronounced influence. At equilibrium the charge susceptibility $\Pi(\omega_p)\approx 0$ (in the window $|\varepsilon|< 0.332$), which is related to
the spectrum of the ABS. There is a pair of the ABS at $E_{\pm}^{\text{eh}}=(\delta\pm\Omega)/2$ for the (e$\uparrow$,h$\downarrow$) channel and another one  $E_{\pm}^{\text{he}}=(-\delta\pm\Omega)/2$ for the
(h$\downarrow$,e$\uparrow$) channel.  In the presented case both pairs of the ABS are beyond the transport window for $|\varepsilon|< \sqrt{\delta^2-\gamma_S^2} = 0.332$; therefore, $\Pi(\omega_p)$ is  exponentially small and the cavity is not disturbed by the nanosystem. For
the bias $|e|V/h =$ 0.4, $|S_{11}|$ exhibits pronounced changes around the resonance at $\varepsilon_r/h= 1.09087$ [see the resonance at the right hand side in  Fig.\ref{mapad6}(b)]. In this region, $\Pi(\omega_p)$ is a complicated function governed by various transitions between the ABS.
For the large bias, $V\rightarrow\infty$, one gets the simple analytical form of the charge susceptibility, Eq.(\ref{chi1e1einf}), and the cavity spectrum, Fig.\ref{mapa}(c).

\section{Summary and Final remarks}

In summary, we have simulated the response of the cavity QED coupled to the DQD-CPS  and analysed the spectrum of the local charge
susceptibility $\Pi(\omega_p)$, which exhibits dynamics of the photon-induced coherent electron-hole recombination processes related with transfers between the ABS. The spectrum of $\Pi(\omega_p)$ shows strong fluctuations at the resonant point, $\hbar\omega_r=\pm\Omega$; it is symmetric at equilibrium and
 becomes asymmetric around $\hbar\omega_r=\Omega$ when the bias voltage increases. In the limit $V\rightarrow\infty$ one gets a simple exact analytical formula, Eq.(\ref{chi1e1einf}), which shows dissipation processes related with intra-level charge fluctuations as well as photon-induced transitions between the ABS.
The conditions to observe these features  are optimal for the level detuning $\delta=0$, when the response can be analyzed as a function of $\varepsilon=(\varepsilon_1+\varepsilon_2)/2$.  For a large $\delta$, the response spectrum becomes very complex and some effects are spoiled.

We have considered the simple model of DQD-CPS restricted the Hilbert space to the sector comprising the DQD in the inter-dot singlet pairing, to get a simple picture of the Andreev bound states in the subgap region.
A key feature of the model is the factorization of the Green functions, Eq.(\ref{eq:greentot}), which corresponds to perfect entanglement of the split Cooper pairs and the separation of the crossed  Andreev  reflections (CAR) for an electron-hole (e$\uparrow$,h$\downarrow$) and a hole-electron (h$\downarrow$,e$\uparrow$) scattering channels.  If direct inter-dot electron hopping is relevant (as for CNT-DQD-CPS in Refs.[\onlinecite{Bruhat2018}] and [\onlinecite{Chevallier2011}]), the condition (\ref{eq:greentot}) is broken, both scattering channels are correlated and the splitter efficiency is reduced.

The superconducting proximity effect is fundamental for the formation of ABS and the operation of CPS. This effect is quantified by parameter $\gamma_S$, which in our calculations has been taken as $\gamma_S/h=0.5$ GHz.
In the experiment on CNT-DQD-CPS~\cite{Bruhat2018} this parameter was estimated as 0.4 GHz,
and it was treated  as a small expansion parameter in comparison to  inter-dot electron hopping $t_b/h= 6.3$ GHz.
The other experiment~\cite{Gramich2017}, on a single proximized CNT quantum dot, showed that the coupling can be $\gamma_S/h=28\div 42$ GHZ (or even much larger). The strong proximity effect was observed in many other quantum dot systems, for example in InAS quantum dots in the recent experiment~\cite{Scherubl2022} (where $\gamma_S/h=35$ GHz).

In our research the strong cavity coupling has been assumed
($\Pi\approx g^2/\gamma_S\gg\kappa$),
to mimic the experimental setup where the analysis of internal dynamics of the nanosystem would be possible.
 The coupling parameter has been set to $g/h=0.4$ GHz close to the experimental value for a triple quantum dot qubit~\cite{Kratochwil2021}. We have considered the single-sided cavity configuration and analyzed the reflectance spectrum $|S_{11}|$, however, one can get easily the cavity transmission for the two-sided resonator configuration with symmetric mirrors~\cite{Walls2008,Cottet2017}
\begin{eqnarray}
S_{12}\equiv\frac{b_t}{b_{in}}=\frac{-\imath \kappa_{ext}}{\omega_{p}-\omega_r-\imath (\kappa_{int}+2\kappa_{ext})/2-\Pi(\omega_{p})}.
\end{eqnarray}

We have also assumed that the resonator is coupled only to the 1st QD, however, both QDs could be coupled to photons, as for example in Ref.~[\onlinecite{Bruhat2018}]. In such a case an inter-dot charge susceptibility should be taken into account. In the limit  $V\rightarrow\infty$ we get $\chi_{1\text{e}2\text{h}}(\omega_p)=-\chi_{1\text{e}1\text{e}}(\omega_p)$, which means that  for a symmetric coupling, $g_1=g_2$, the total charge susceptibility $\Pi(\omega_p)=0$ (as one could expect from the current conservation rule).  This resembles the situation for the DQD with normal metallic contacts, where an asymmetric coupling of two dots to the cavity is required~\cite{Burkard2020,Cottet2017}.
For a finite bias, $\Pi(\omega_p)$ becomes finite but small,  due to displacement currents which should be taken into account.

We hope that the paper will inspire experimentalists to perform a direct studies into the spectrum of the Andreev bound states using microwave cavity spectroscopy.
\bibliographystyle{apsrev4-2}
\bibliography{cav-DQD-CPS}

\end{document}